\documentclass[runningheads]{llncs}
\usepackage[T1]{fontenc}
\usepackage[utf8]{inputenc}
\usepackage{graphicx}
\usepackage{amsmath,amssymb}
\usepackage{booktabs}
\usepackage{hyperref}
\usepackage{cleveref}
\usepackage{xcolor}
\usepackage{subcaption}
\usepackage{float}
\usepackage{placeins}
\usepackage{tikz}
\usepackage{pgfplots}
\pgfplotsset{compat=1.17}
\usetikzlibrary{arrows.meta,positioning}
\usepackage{adjustbox}
\usepackage{multirow}
\usepackage{enumitem}
\usepackage{orcidlink}
\setlist{nosep,leftmargin=*}

\begin{document}

\title{A Treasure Trove of Performance: Analyzing the IO500 Submission Data}

\author{Julian Kunkel\,\orcidlink{0000-0002-6915-1179} \and
Aasish Kumar Sharma\,\orcidlink{0000-0002-7514-2340} \and
Anila Ghazanfar\,\orcidlink{0009-0003-2894-6893} \and
Sepehr Mahmoodianhamedani\,\orcidlink{0009-0000-0569-8047} \and
Sascha Safenreider\,\orcidlink{0009-0005-2287-0060}}

\authorrunning{J. Kunkel et al.}

\institute{Georg-August-Universit\"at G\"ottingen / GWDG mbH, G\"ottingen, Germany\\
\email{\{julian.kunkel, aasish-kumar.sharma, anila.ghazanfar, s.mahmoodian, sascha.safenreider\}@gwdg.de}}

\maketitle

\begin{abstract}
\emergencystretch=3em
The IO500 benchmark has become the community standard for evaluating high-performance computing (HPC) storage system performance, yet the detailed data contained in its submission packages remains largely unexplored beyond aggregate leaderboard rankings.
We present a statistical characterization of 61~IO500 submissions from four competition lists (ISC21 through SC22), examining score distributions, inter-phase correlations, and insights derived from detailed log files that accompany each submission.
Our analysis reveals that IO500 scores span four orders of magnitude.
Spearman correlation analysis shows strong within-domain clustering for both bandwidth ($r_s = 0.78$\,to\,$0.96$) and metadata ($r_s = 0.89$\,to\,$0.98$) phases, with the composite sub-scores exhibiting $r_s = 0.92$ at per-node level (Pearson $r = 0.53$).
Log-level analysis uncovers file-system-specific patterns in Interleaved Or Random (IOR) close-time overhead, straggler behavior during the stonewall wear-down phase, and parallel-find load imbalance that are invisible in aggregate scores.
These findings demonstrate that IO500 submission packages constitute a valuable research resource for understanding storage system behavior; data and scripts are described in \Cref{sec:data}.
\end{abstract}


\section{Introduction}\label{sec:intro}

Benchmarking storage systems is central to procurement decisions, deployment validation, and performance understanding in high-performance computing (HPC).
The IO500~\cite{kunkel2016establishing,monnier2019profiling} has become the community standard for this purpose, combining bandwidth measurements - Interleaved Or Random (IOR) with metadata evaluations - MetaData Test (MDTest) into a composite score.
Since its establishment, hundreds of submissions have been collected, each containing aggregate scores and detailed log files with per-process timing, stonewall statistics, and architectural metadata.

Despite this wealth of data, published analyses have focused primarily on leaderboard rankings and top-performing systems~\cite{kunkel2019tracking}.
The detailed submission packages constitute a largely untapped research resource.
Systematic analysis can reveal patterns in storage system behavior, relationships between benchmark phases, and performance characteristics that aggregate scores obscure.

We present a descriptive statistical characterization of 61~IO500 submissions from four competition lists spanning 2021 to 2022 (ISC21, SC21, ISC22, SC22).
Our contributions are:
\begin{enumerate}
    \item \textbf{Descriptive characterization} of IO500 score distributions and variability across file systems, interconnects, and deployment scales.
    \item \textbf{Correlation analysis} between benchmark phases using Spearman rank correlations with Benjamini-Hochberg's False Discovery Rate (FDR) correction, identifying distinct bandwidth and metadata clusters.
    \item \textbf{Log-derived insights} from per-process timing data, revealing file-system-specific patterns in close-time overhead, straggler behavior, and parallel-find load balancing.
\end{enumerate}

We emphasize that our findings describe patterns in this specific dataset and should be interpreted cautiously when generalizing.
All analysis scripts and the full IO500 submission data are publicly available (see~\Cref{sec:methodology}).

\section{Background: The IO500 Benchmark}\label{sec:background}

The IO500 benchmark suite~\cite{kunkel2016establishing} characterizes storage performance through two primary components and a find phase.

\paragraph{Interleaved Or Random (IOR)~\cite{shan2007using}:} It measures bandwidth in two configurations.
\emph{IOR-easy} uses a file-per-process layout with large sequential I/O, emulating optimized scientific applications.
\emph{IOR-hard} uses a single shared file with 47{,}008-byte interleaved records, stressing non-contiguous small-transfer access typical of checkpoint/restart workloads.
Both report throughput in GiB/s.

\paragraph{MetaData Test (MDTest)~\cite{mdtest2007}:} It measures metadata operation rates (create, stat, read, delete) in two modes.
\emph{MDTest-easy} uses a directory-per-process layout, where each process creates its own unique directory containing zero-byte files.
\emph{MDTest-hard} operates within a single shared directory with small data payloads.
Results are in kIOPS (thousands of I/O operations per second).

\paragraph{Parallel find (pfind):} It measures namespace traversal rate across the file trees created by MDTest, reported in kIOPS.

\paragraph{Scoring.}
Composite scores use geometric means:
\begin{align}
    \text{Score}_{\text{BW}} &= \bigl(S_{\text{ior-easy-w}} \cdot S_{\text{ior-easy-r}} \cdot S_{\text{ior-hard-w}} \cdot S_{\text{ior-hard-r}}\bigr)^{1/4} \label{eq:scorebw} \\
    \text{Score}_{\text{MD}} &= \bigl(S_{\text{md-easy-w}} \cdot S_{\text{md-easy-s}} \cdot S_{\text{md-hard-w}} \cdot S_{\text{md-hard-s}} \cdot S_{\text{find}}\bigr)^{1/5} \label{eq:scoremd} \\
    \text{Score}_{\text{overall}} &= \sqrt{\text{Score}_{\text{BW}} \times \text{Score}_{\text{MD}}} \label{eq:scoreoverall}
\end{align}
The overall score thus combines fundamentally different units (bandwidth and metadata rate) via geometric aggregation.

\paragraph{Submission packages:} They include per-process Comma-Separated Values (CSV) timing data (start/end times per operation), stonewall throughput (performance at the moment the 300-second time limit is reached), and file-closing durations.
The stonewall mechanism ensures write phases run for at least 300~seconds; afterward, all processes complete in-flight operations in a ``wear-down'' phase, emulating bulk-synchronous behavior and exposing straggler effects.

\section{Related Work}\label{sec:related}

IOR~\cite{shan2007using} and MDTest~\cite{mdtest2007} have long been used individually for HPC storage characterization.
Kunkel et al.~\cite{kunkel2016establishing} unified them into the IO500 with standardized configurations and composite scoring.
Monnier et al.~\cite{monnier2019profiling} combined IO500 with the Mistral profiling tool for platform-level storage evaluation.

I/O variability and workload characterization have received extensive study.
Lofstead et al.~\cite{lofstead2010managing} analyzed variability in petascale storage systems; Luu et al.~\cite{luu2015multiplatform} conducted cross-platform I/O studies.
Carns et al.~\cite{carns2011understanding} and Snyder et al.~\cite{snyder2016modular} developed continuous characterization via Darshan, and Lockwood et al.~\cite{lockwood2017umami} proposed holistic analysis through the UMAMI framework.
Major parallel file systems including Lustre~\cite{schwan2003lustre}, General Parallel File System (GPFS)~\cite{schmuck2002gpfs}, and Distributed Asynchronous Object Storage (DAOS)~\cite{daos2024,hennecke2023understanding} have been studied individually using various benchmarks.

Despite this body of work, no prior study systematically analyzes the IO500 submission corpus itself as a statistical dataset, examining inter-phase relationships and extracting insights from the per-process log files that accompany each submission.
Our work addresses this gap.

\section{Methodology}\label{sec:methodology}

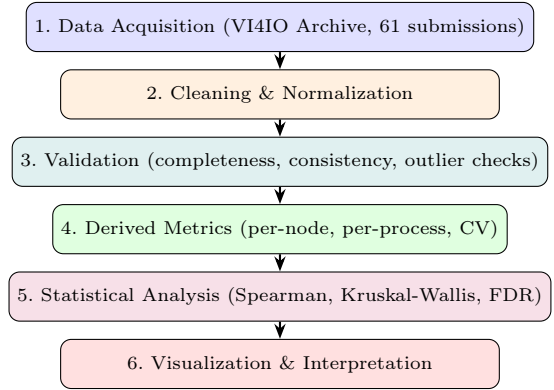
\begin{figure}[t]
\centering
\begin{tikzpicture}[
    node distance=1.0em,
    box/.style={rectangle, draw, rounded corners, minimum width=7.8cm, minimum height=0.65cm, text centered, font=\scriptsize, fill=#1!12},
    arrow/.style={-{Stealth[length=2mm]}, thick}
]
\node[box=blue] (acq) {1.\ Data Acquisition (VI4IO Archive, 61 submissions)};
\node[box=orange, below=of acq] (clean) {2.\ Cleaning \& Normalization};
\node[box=teal, below=of clean] (valid) {3.\ Validation (completeness, consistency, outlier checks)};
\node[box=green, below=of valid] (derive) {4.\ Derived Metrics (per-node, per-process, CV)};
\node[box=purple, below=of derive] (stat) {5.\ Statistical Analysis (Spearman, Kruskal-Wallis, FDR)};
\node[box=red, below=of stat] (viz) {6.\ Visualization \& Interpretation};
\draw[arrow] (acq) -- (clean);
\draw[arrow] (clean) -- (valid);
\draw[arrow] (valid) -- (derive);
\draw[arrow] (derive) -- (stat);
\draw[arrow] (stat) -- (viz);
\end{tikzpicture}
\caption{Analysis pipeline for IO500 repository characterization.}\label{fig:pipeline}
\end{figure}

Our analysis pipeline (\Cref{fig:pipeline}) proceeds from data acquisition through statistical analysis and visualization.

\subsection{Data Acquisition and Scope}\label{sec:data}

We analyze 61~IO500 submissions obtained from Virtual Institute for I/O (VI4IO) archive~\cite{vi4io2023} snapshots, spanning four competition lists: ISC21 (June~2021), SC21 (November~2021), ISC22 (June~2022), and SC22 (November~2022).
The distribution is 24~submissions from 2021 lists and 37 from 2022 lists.
These submissions were selected because their full packages (including log files) were available at data collection time; they represent the complete set of submissions for which detailed per-process timing data could be retrieved, not a curated sample.
The dataset spans five file system families (Lustre, GPFS/Spectrum Scale, DAOS, WekaFS, BeeGFS) and three major interconnect tiers, but vendor representation is uneven: Lustre accounts for 44\% of submissions, and multiple submissions often originate from variants of the same vendor platform.
Conclusions should therefore be interpreted as characterising this particular corpus rather than the broader space of HPC storage systems.

Since our initial analysis, the IO500 community has released the full submission dataset on GitHub~\cite{io500submission2025}, now containing data from 131+~sites.
We use the original 61-submission subset; extending to the full repository is future work (\Cref{sec:conclusion}).
Our analysis scripts are at \url{https://gitlab-ce.gwdg.de/hpc-team/io500-analysis}.

\subsection{Data Cleaning and Quality}\label{sec:cleaning}

Cleaning steps include: file system name standardization (e.g., grouping GPFS and Spectrum Scale), interconnect speed normalization to Gb/s, and exclusion of records with missing values on a per-analysis basis.

\emph{Data quality limitations:}
Self-reported submission metadata contains known quality issues.
Interconnect speed is sometimes ambiguous: some submissions report throughput exceeding nominal single Network Interface Card (NIC) speed, likely because NIC count per node is not consistently reported.
We treat interconnect speed as a categorical grouping variable rather than a precise measurement and document these inconsistencies rather than attempting correction.

\subsection{Derived Metrics and Statistical Methods}\label{sec:stats}
In addition to official IO500 scores, several derived metrics provide deeper insights. Per-node performance was calculated as:
\begin{equation}
\text{Score}_{\text{per-node}} = \frac{\text{Overall Score}}{\text{Number of Client Nodes}}
\end{equation}
This metric removes scale effects, allowing direct comparison of architectural efficiency across deployments of different sizes. Similarly, per-process performance was defined as:
\begin{equation}
\text{Score}_{\text{per-proc}} = \frac{\text{Overall Score}}{\text{Nodes} \times \text{Processes per Node}}
\end{equation}
The coefficient of variation (CV) quantifies variability for each metric across submissions:
\begin{equation}
\text{CV} = \frac{\sigma}{\mu} \times 100\%
\end{equation}
where $\sigma$ denotes standard deviation and $\mu$ the mean. High CV values indicate inconsistent performance, an important factor for production system predictability.

To compare systems of different scales, we compute per-node and per-process normalized scores.
Per-node normalization removes the dominant effect of deployment scale, enabling architectural efficiency comparison.
We note that dividing the composite overall score (a geometric mean spanning GiB/s and kIOPS) by node count produces a mixed-unit ratio; individual phase normalizations are more directly interpretable.

\emph{Correlation analysis:}
We use Spearman rank correlation ($r_s$) as our primary measure, appropriate for the skewed, non-normal distributions typical of benchmark data.
Pearson correlations are reported for comparison.
Multiple comparisons across the correlation matrix are controlled via Benjamini-Hochberg's FDR correction~\cite{benjamini1995controlling} at $\alpha = 0.05$.

\emph{Group comparisons:}
Kruskal-Wallis H-tests~\cite{kruskal1952use} compare per-node performance across interconnect speed groups, with eta-squared ($\eta^2 = H/(n{-}1)$) effect sizes reported alongside $p$-values.
The independence assumption may be violated when multiple submissions originate from the same organization; reported $p$-values are therefore approximate.

\section{Dataset Characteristics}\label{sec:dataset}

\Cref{tab:dataset} summarizes the composition of our dataset.
Lustre dominates with 27 submissions (44\%), followed by GPFS/Spectrum Scale (12), DAOS~(10), WekaFS~(7), and BeeGFS/others~(5).
InfiniBand HDR (200\,Gb/s) and EDR (100\,Gb/s) together account for 40 of 61~submissions.
Public IO500 submissions likely over-represent well-tuned configurations.
The 2021-2022 window captures a period of rapid evolution, particularly the emergence of DAOS.

\begin{table}[t]
\centering
\caption{Dataset composition: 61 IO500 submissions by file system and interconnect.}\label{tab:dataset}
\begin{adjustbox}{max width=\textwidth}
\begin{tabular}{lr@{\qquad}lr}
\toprule
\textbf{File System} & \textbf{N} & \textbf{Interconnect} & \textbf{N} \\
\midrule
Lustre              & 27 & IB HDR (200\,Gb/s)     & 22 \\
GPFS/SpectrumScale  & 12 & IB EDR (100\,Gb/s)     & 18 \\
DAOS                & 10 & Omni-Path (100\,Gb/s)  & 11 \\
WekaFS              &  7 & Other / Unknown        & 10 \\
BeeGFS / Other      &  5 &                        &    \\
\bottomrule
\end{tabular}
\end{adjustbox}
\end{table}

\section{Results}\label{sec:results}

\subsection{Score Distributions}\label{sec:distributions}

\begin{figure}[t]
    \centering
    \includegraphics[width=0.85\textwidth]{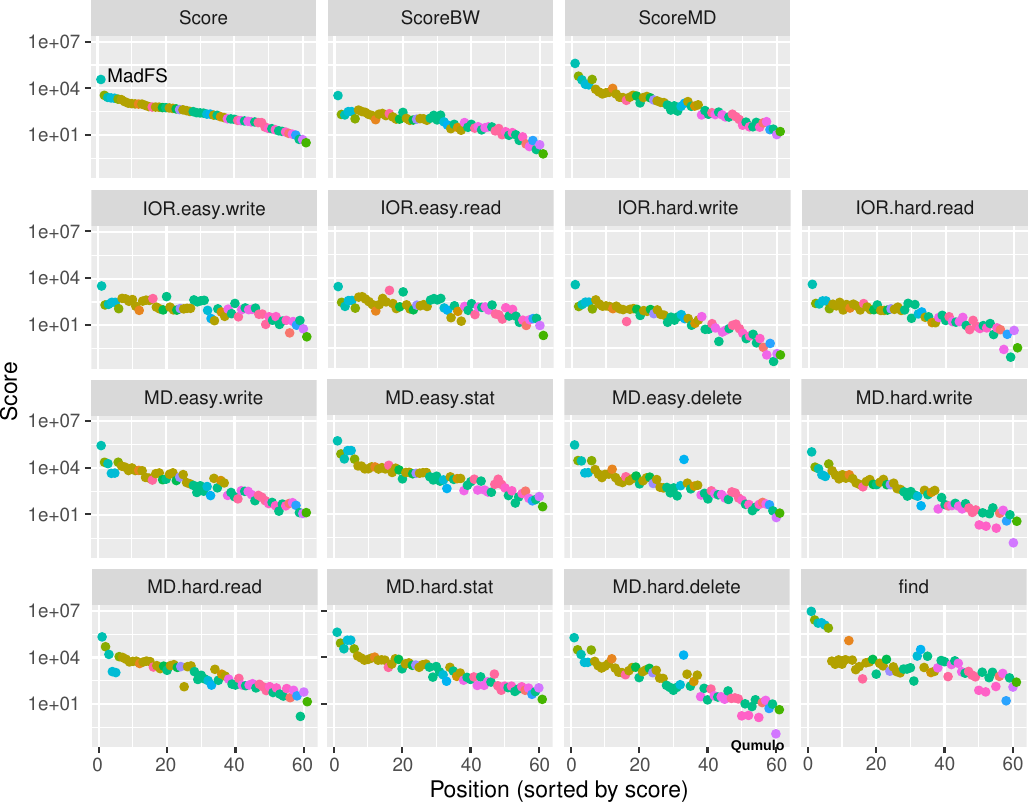}\\[2pt]
    \includegraphics[width=0.85\textwidth]{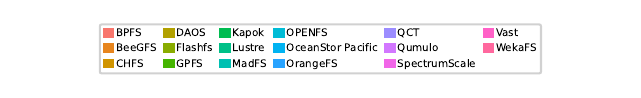}
    \caption{IO500 scores across 61~submissions, colored by file system type (log scale). Scores span approximately four orders of magnitude, reflecting substantial heterogeneity in system scale and architecture.}\label{fig:scores}
\end{figure}

\Cref{fig:scores} shows overall IO500 scores spanning approximately four orders of magnitude.
\Cref{tab:summary} presents summary statistics; all metrics exhibit coefficients of variation of at least~2.0, with ``hard'' configurations (IOR-hard, MDTest-hard) consistently showing higher variability than their ``easy'' counterparts.
This indicates that the more challenging I/O and metadata access patterns are more sensitive to architectural differences than the corresponding easy configurations.
The mean substantially exceeds the median in all cases, confirming right-skewed distributions driven by a few high-performance systems.

\begin{table}[t]
\centering
\caption{Summary statistics for IO500 metrics across 61~submissions.}\label{tab:summary}
\begin{adjustbox}{max width=\textwidth}
\begin{tabular}{lrrrrr}
\toprule
\textbf{Metric} & \textbf{Min} & \textbf{Median} & \textbf{Mean} & \textbf{Max} & \textbf{CV} \\
\midrule
Overall Score        & 3.2   & 254   & 1{,}128   & 36{,}850   & 4.17 \\
Score BW (GiB/s)     & 0.6   & 88    & 156       & 3{,}422    & 2.79 \\
Score MD (kIOPS)     & 10.5  & 837   & 10{,}641  & 396{,}873  & 4.82 \\
IOR-easy Write       & 1.8   & 113   & 220       & 3{,}360    & 2.00 \\
IOR-hard Write       & 0.0   & 24.8  & 122       & 3{,}717    & 3.91 \\
MDTest-easy Create   & 12.0  & 768   & 7{,}639   & 278{,}101  & 4.66 \\
MDTest-hard Create   & 0.1   & 279   & 3{,}016   & 102{,}419  & 4.40 \\
\bottomrule
\end{tabular}
\end{adjustbox}
\end{table}

\subsection{Correlation Analysis}\label{sec:correlations}

\begin{figure}[t]
    \centering
    \begin{subfigure}{0.48\textwidth}
        \centering
        \includegraphics[width=\textwidth]{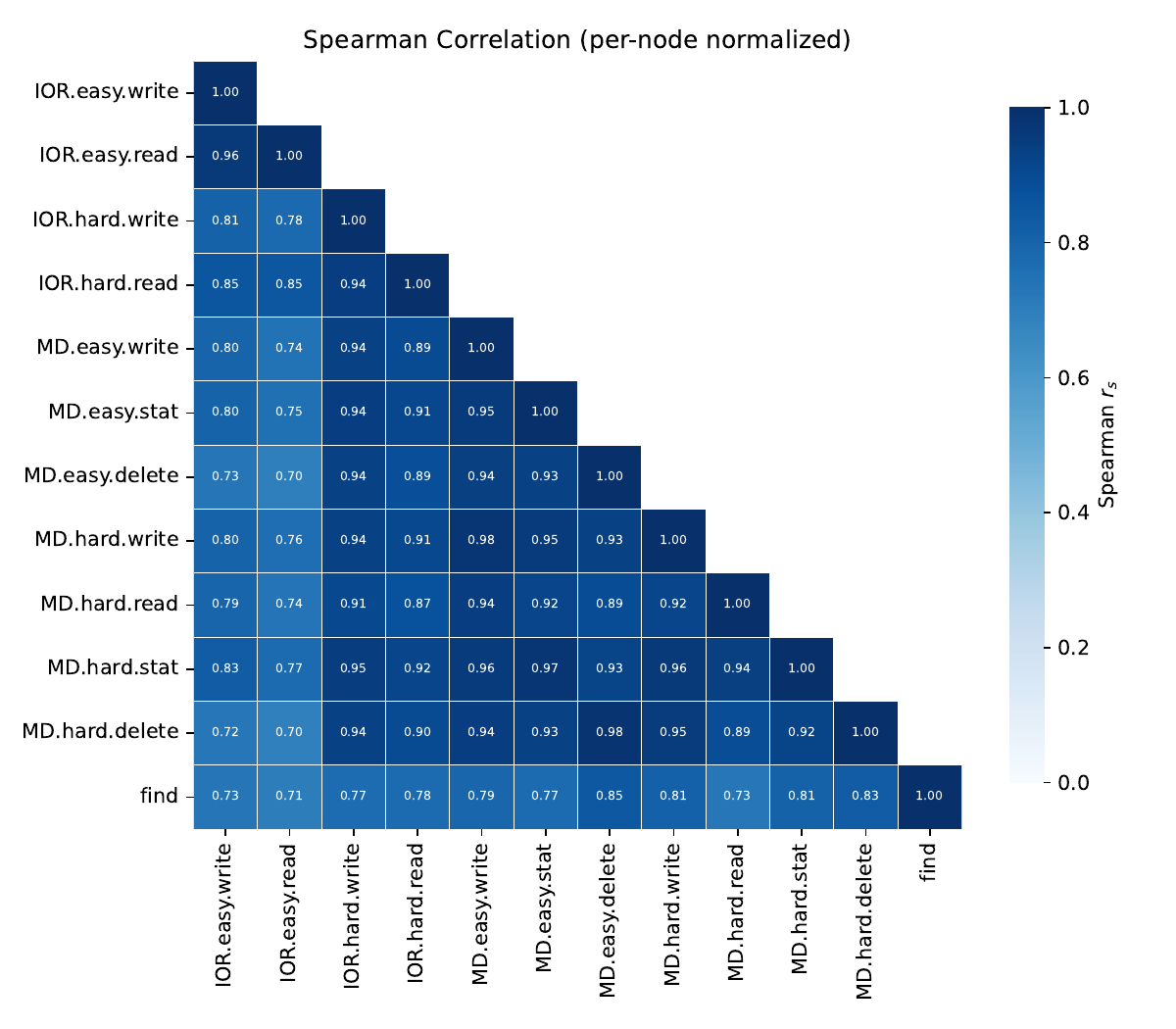}
        \caption{Per-node normalized}\label{fig:corr-pernode}
    \end{subfigure}
    \hfill
    \begin{subfigure}{0.48\textwidth}
        \centering
        \includegraphics[width=\textwidth]{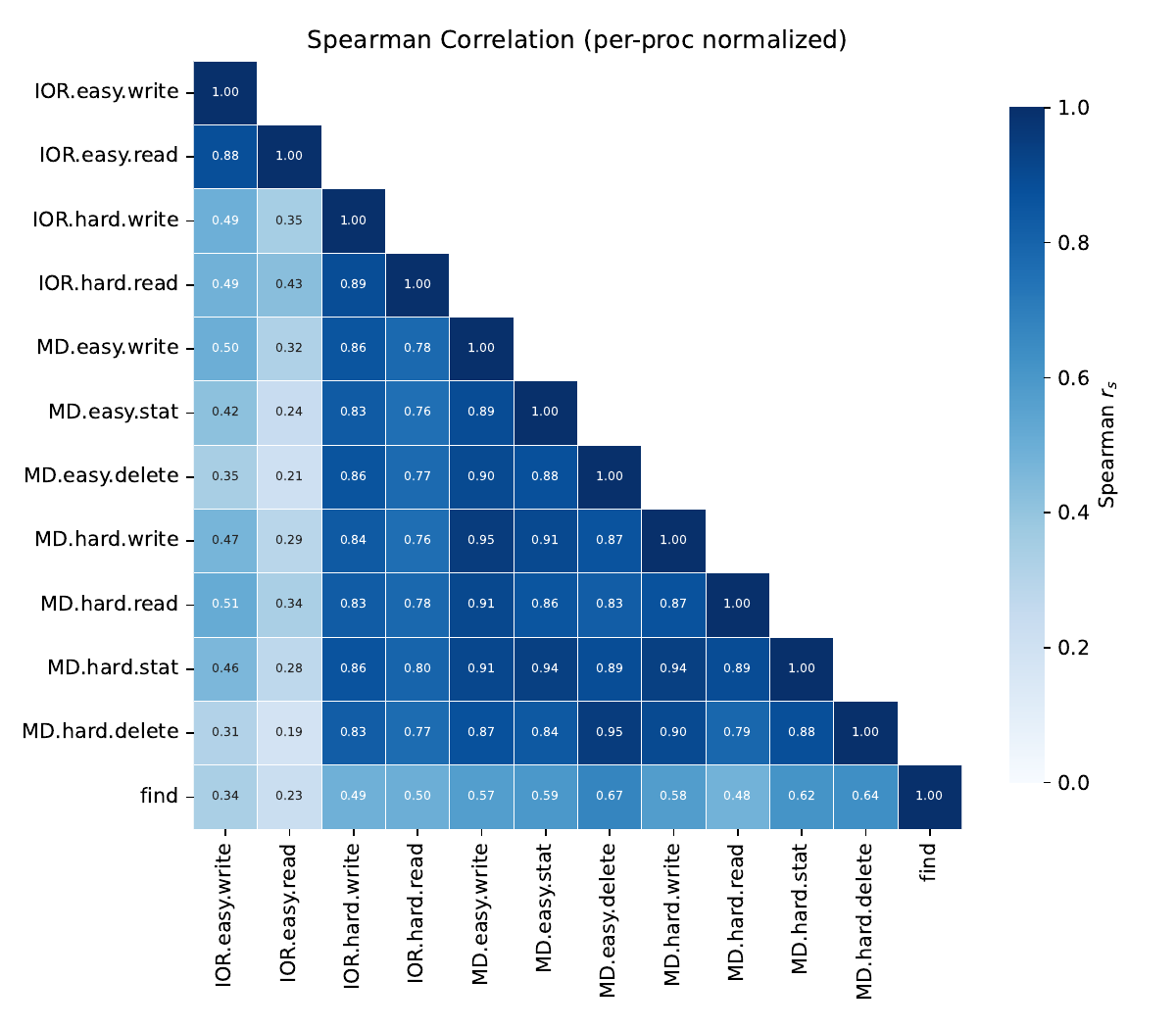}
        \caption{Per-process normalized}\label{fig:corr-perproc}
    \end{subfigure}
    \caption{Spearman rank correlation matrices for IO500 phase scores.
    The box color intensity encodes correlation magnitude.
    All displayed correlations are significant after Benjamini-Hochberg's FDR correction at $\alpha = 0.05$.}\label{fig:correlations}
\end{figure}

\Cref{fig:correlations} presents Spearman rank correlation matrices for per-node and per-process normalized scores.
Two distinct clusters emerge: IOR bandwidth phases correlate strongly within-domain ($r_s = 0.78$\,to\,$0.96$), as do MDTest metadata phases ($r_s = 0.89$\,to\,$0.98$).
Cross-domain correlations between IOR and MDTest phases, while still positive, are generally weaker in the per-process normalized view, supporting the IO500's multi-dimensional scoring design.

The composite sub-scores Score\textsubscript{BW} and Score\textsubscript{MD} exhibit a strong correlation ($r_s = 0.92$, Pearson $r = 0.53$) at the per-node level, despite the weaker component-level cross-domain correlations.
This is a consequence of geometric mean aggregation combined with residual scale effects: systems with more resources tend to achieve higher scores in both domains.

Pearson correlations show consistent patterns with somewhat different magnitudes due to outlier influence.
The rank-based Spearman results are more robust to the skewed distributions observed above and are our primary measure.

\paragraph{Interpretation:}
These correlation patterns suggest that some benchmark phases capture partially overlapping performance characteristics \emph{in this dataset}.
However, we caution against interpreting observed correlations as evidence that any phase is redundant in general.
The IO500 is designed to cover a broad space of possible storage system behaviors; correlation patterns in 61~well-tuned systems may not hold for emerging architectures or differently configured deployments.
Design decisions about benchmark composition should consider the full range of intended use cases, not correlation patterns from a single sample.

\subsection{Interconnect Speed and Per-Node Performance}\label{sec:interconnect}

\begin{figure}[t]
    \centering
    \begin{subfigure}{0.32\textwidth}
        \centering
        \includegraphics[width=\textwidth]{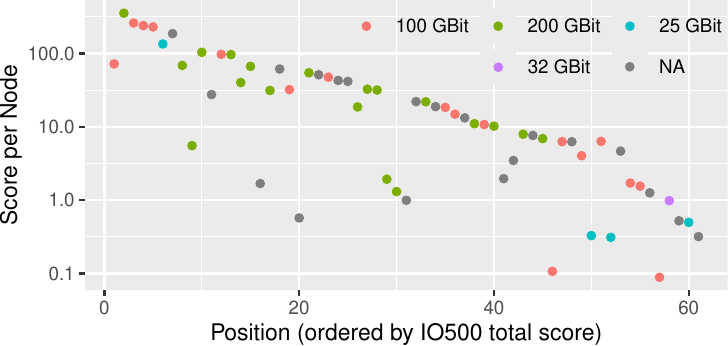}
        \caption{Overall score}\label{fig:intercon-overall}
    \end{subfigure}
    \hfill
    \begin{subfigure}{0.32\textwidth}
        \centering
        \includegraphics[width=\textwidth]{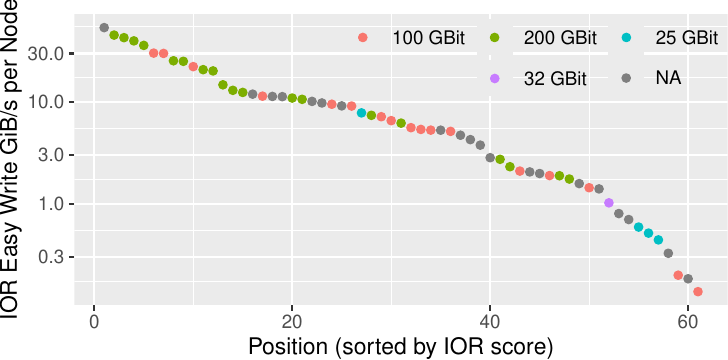}
        \caption{IOR-easy write}\label{fig:intercon-ior}
    \end{subfigure}
    \hfill
    \begin{subfigure}{0.32\textwidth}
        \centering
        \includegraphics[width=\textwidth]{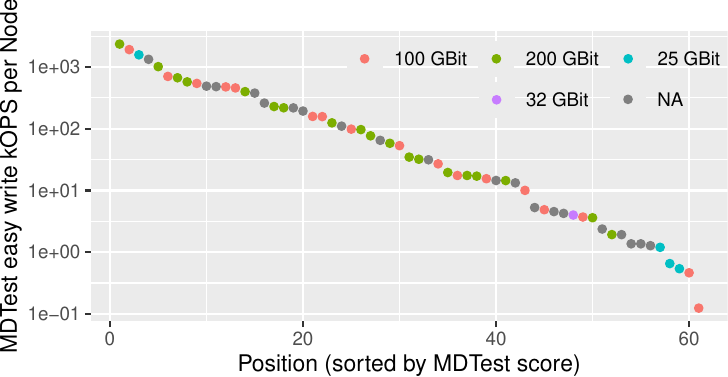}
        \caption{MDTest-easy write}\label{fig:intercon-mdtest}
    \end{subfigure}
    \caption{Per-node IO500 performance grouped by reported interconnect speed.
    Higher speeds are \emph{associated with} higher per-node performance, with substantial within-group variability.}\label{fig:interconnect}
\end{figure}

\Cref{fig:interconnect} shows per-node scores grouped by interconnect speed.
Higher speeds are generally associated with higher per-node performance.
Kruskal-Wallis tests yield $H = 4.40$, $p = 0.111$, $\eta^2 = 0.11$ for overall per-node score, indicating a moderate effect size but falling short of statistical significance at $\alpha = 0.05$.
For IOR-easy per-node bandwidth, the association is stronger ($H = 11.16$, $p = 0.004$, $\eta^2 = 0.28$), consistent with the expectation that large sequential I/O is more directly constrained by network bandwidth than metadata operations.

We use associational language deliberately.
Confounding factors including storage media type, file system architecture, storage server count, and tuning quality vary simultaneously across submissions.
Group sizes are unequal (as few as $n = 5$), limiting statistical power.
Reported interconnect speeds may not reflect effective per-node bandwidth due to unreported NIC counts (\Cref{sec:cleaning}).
These limitations prevent causal conclusions about the role of interconnect speed.

\section{Log-Derived Insights}\label{sec:logs}

Beyond aggregate scores, IO500 submission packages contain detailed log files that reveal performance characteristics invisible in summary metrics.

\subsection{Runtime Distributions and Close-Time Overhead}\label{sec:runtime}

\begin{figure}[!t]
    \centering
    \includegraphics[width=0.85\textwidth]{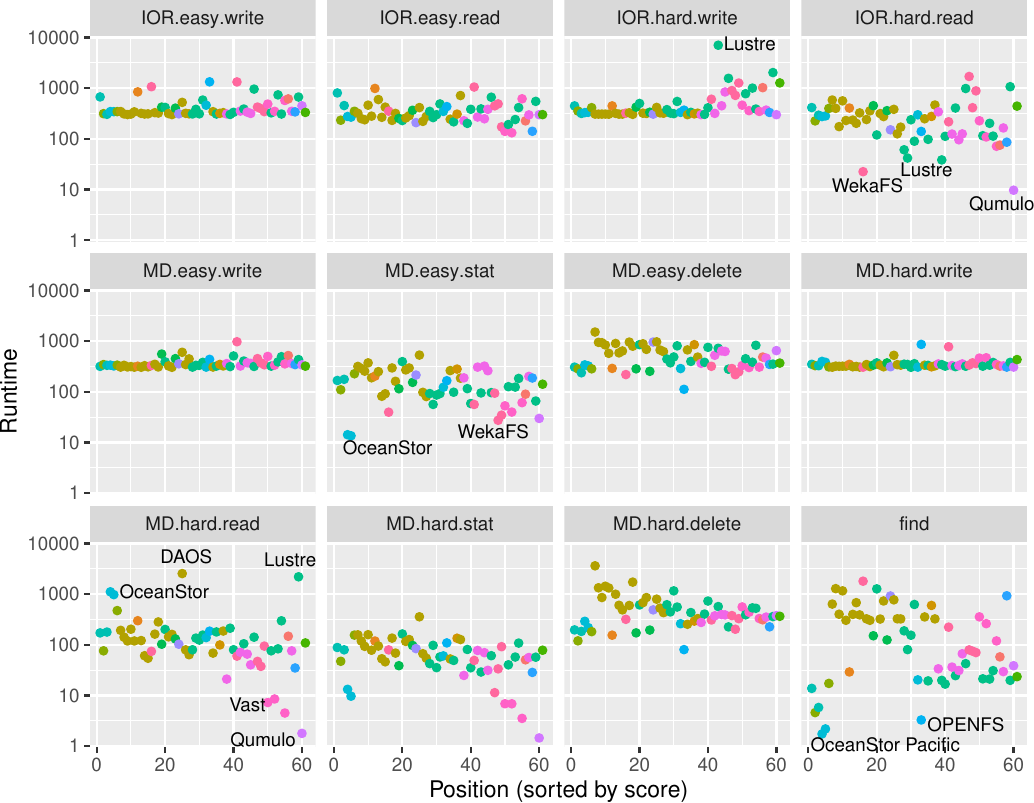}\\[2pt]
    \includegraphics[width=0.85\textwidth]{plots/overall-score-legend.pdf}
    \caption{Runtime distributions across IO500 phases.
    Write phases are bounded below by the 300-second stonewall; some read/stat phases complete in under 10~seconds (possible residual caching effects).}\label{fig:runtime}
\end{figure}

\Cref{fig:runtime} shows phase runtimes.
Write phases consistently meet or exceed the 300-second stonewall, with some extending beyond 3{,}600~seconds during the wear-down phase.
Some read and stat phases complete in approximately 10~seconds or less, raising questions about residual caching despite IO500's read-after-write rules.
We flag these submissions as potentially cache-affected without excluding them.

\begin{figure}[!t]
    \centering
    \includegraphics[width=0.85\textwidth]{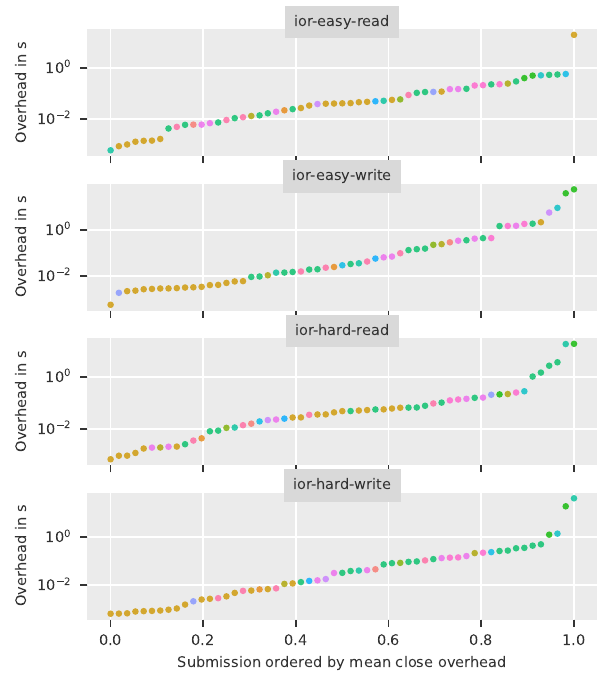}\\[2pt]
    \includegraphics[width=0.85\textwidth]{plots/overall-score-legend.pdf}
    \caption{IOR close-time overhead by file system.
    Lustre shows up to tens of seconds (cache flush and metadata finalization); DAOS shows negligible overhead.
    Close time is included in the IOR timing measurement.}\label{fig:close}
\end{figure}

A particularly valuable piece of hidden information in IOR log files is the time spent in the file close operation.
\Cref{fig:close} shows close-time overhead by file system.
Lustre systems exhibit close times of up to tens of seconds, reflecting the cost of flushing data from client caches and completing metadata operations.
DAOS shows negligible close overhead, consistent with its persistent-memory-based architecture~\cite{daos2024,hennecke2023understanding}.

This observation is practically important: close time is included in the IOR timing measurement and represents genuine application-visible I/O cost.
The variation across file systems highlights that different architectures distribute the cost of data persistence differently between write and close phases.

\subsection{Stonewall-Relative Analysis}\label{sec:stonewall}

\begin{figure}[!t]
    \centering
    \begin{subfigure}{0.48\textwidth}
        \centering
        \includegraphics[width=\textwidth]{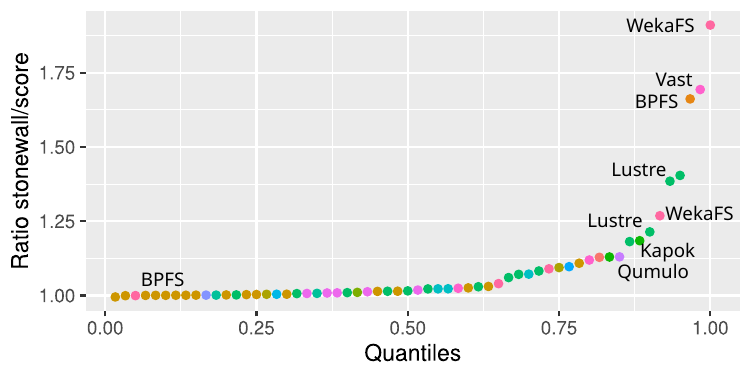}
        \caption{IOR-easy write}\label{fig:qq-ior-easy}
    \end{subfigure}
    \hfill
    \begin{subfigure}{0.48\textwidth}
        \centering
        \includegraphics[width=\textwidth]{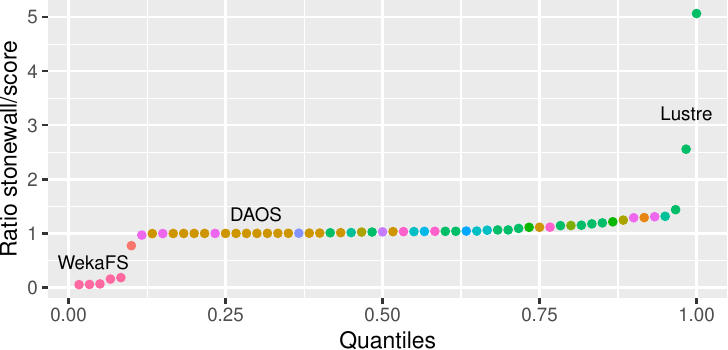}
        \caption{IOR-hard write}\label{fig:qq-ior-hard}
    \end{subfigure}\\[2pt]
    \includegraphics[width=\textwidth]{plots/overall-score-legend.pdf}
    \caption{Stonewall-relative Quantile-Quantile (Q-Q) plots for IOR writes.
    Values near~1.0 indicate uniform process completion; larger values reveal stragglers in the wear-down phase.
    IOR-hard shows substantially larger deviations (up to 2$\times$\,to\,5$\times$).}\label{fig:qq-ior}
\end{figure}

To understand per-process behavior during the stonewall phase, we compute the ratio of each process's total runtime to the stonewall duration.
\Cref{fig:qq-ior} shows these ratios as Q-Q plots for IOR write phases.

IOR-easy processes finish near the stonewall time (ratio $\approx 1.0$) with moderate tail deviations, as expected for the file-per-process pattern where each process writes independently with minimal contention.
IOR-hard shows substantially larger deviations (up to 2$\times$\,to\,5$\times$), reflecting the straggler sensitivity of the shared-file access pattern where all processes contend for the same file.
MDTest phases exhibit analogous patterns: MDTest-hard with larger deviations than MDTest-easy.

File-system-specific patterns are visible.
In particular, some Lustre submissions on Hard Disk Drive (HDD)-based storage show the largest stonewall-relative deviations, while WekaFS submissions display characteristic step-like patterns attributable to internal storage bucket oversubscription.

\subsection{Process-Level Straggler Patterns}\label{sec:stragglers}

\begin{figure}[!t]
    \centering
    \begin{subfigure}{0.48\textwidth}
        \centering
        \includegraphics[width=\textwidth]{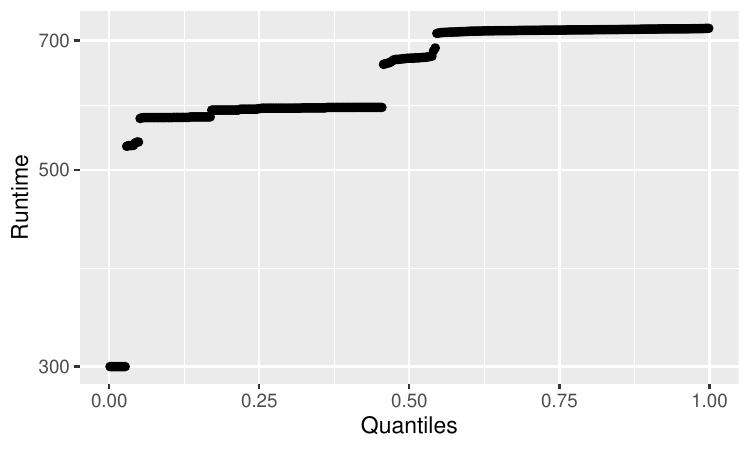}
        \caption{WekaFS: clustered stragglers}\label{fig:straggler-weka}
    \end{subfigure}
    \hfill
    \begin{subfigure}{0.48\textwidth}
        \centering
        \includegraphics[width=\textwidth]{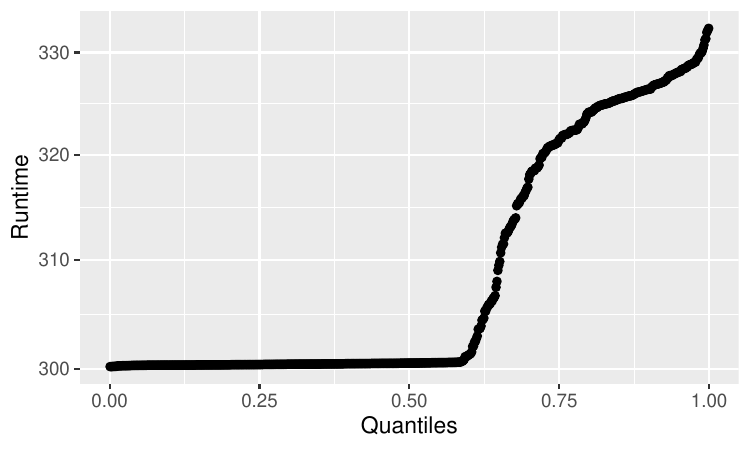}
        \caption{Lustre: contiguous stragglers}\label{fig:straggler-lustre}
    \end{subfigure}
    \caption{Per-process Q-Q plots for IOR-hard write from individual submissions, illustrating contrasting straggler architectures.
    (a)~WekaFS: clustered groups suggest storage-bucket-level contention.
    (b)~Lustre: contiguous slow ranks suggest Object Storage Target (OST)-level hotspots.}\label{fig:stragglers}
\end{figure}

Individual submissions reveal architecture-specific straggler patterns (\Cref{fig:stragglers}).
The WekaFS submission shows \emph{clustered} groups of slow processes, consistent with its distributed architecture where processes mapped to the same storage bucket experience correlated contention.
The Lustre submission shows a \emph{contiguous} range of slow Message Passing Interface (MPI) ranks, consistent with Lustre's OST striping model where a single overloaded storage target affects all mapped processes.

These contrasting patterns demonstrate that process-level log analysis provides architectural insights entirely invisible in aggregate scores.
The straggler structure reveals how each file system distributes load and where contention bottlenecks arise.

\subsection{Parallel Find Load Imbalance}\label{sec:find}

The pfind phase reveals extreme load imbalance: a single process may check over 5~million files while others check approximately 100{,}000.
This arises from the inherent difficulty of parallelizing directory tree traversal, compounded by MDTest-easy's per-process directory structure which cannot be efficiently distributed across find processes.
Job-stealing mechanisms partially mitigate the skew but do not eliminate it.
Runtime metrics (time spent in job stealing, active utilization) reveal that most processes spend the majority of their time waiting rather than actively traversing, indicating opportunities for improved scheduling in the find implementation.

\section{Practical Implications}\label{sec:implications}

Our analysis yields actionable insights for several stakeholder groups:

\paragraph{System procurement:}
The composite IO500 score obscures trade-offs between bandwidth and metadata performance.
Per-node analysis reveals substantial architectural efficiency differences hidden when raw scores (dominated by node count) are compared.
Evaluators should examine per-node scores alongside aggregate rankings and prioritize the phase scores most relevant to their workload profile.

\paragraph{Benchmarking practice:}
Log-level analysis exposes bottlenecks (close-time overhead, straggler behavior) invisible in aggregate throughput numbers.
Stonewall-relative analysis provides a standardized way to assess process-level performance uniformity.
We recommend that benchmark consumers inspect timing breakdowns, not just headline throughput.

\paragraph{IO500 community:}
Submission packages are a valuable research resource beyond rankings.
The public GitHub dataset~\cite{io500submission2025} enables longitudinal studies and cross-site modeling.
We encourage submitters to provide accurate metadata (particularly NIC count and storage target count) to improve cross-submission analysis quality.

\paragraph{Storage system design:}
The file-system-specific straggler patterns (\Cref{sec:stragglers}) provide actionable information about load balancing behavior.
Clustered patterns (WekaFS) and contiguous patterns (Lustre) point to different architectural bottlenecks and may guide optimization efforts for storage target mapping and load distribution.

\section{Conclusion and Future Work}\label{sec:conclusion}

We have presented a statistical characterization of 61~IO500 submissions from the 2021-2022 competition lists.
Our analysis documents substantial score variability (CV~$\geq$~2 across all metrics), identifies distinct bandwidth and metadata correlation clusters ($r_s = 0.78$\,to\,$0.98$ within, $r_s = 0.70$\,to\,$0.95$ across domains), and demonstrates the value of log-level analysis for uncovering file-system-specific performance patterns in close-time overhead, straggler behavior, and find load imbalance.

\emph{Limitations:}
Our 61-submission dataset is a subset of available data; observed patterns may not generalize.
Self-reported metadata contains quality issues, particularly regarding interconnect specifications.
Statistical tests assume independence between submissions, which is violated when multiple submissions originate from the same site.
Vendor representation is skewed: Lustre alone accounts for 44\% of submissions, and several submissions represent different configurations of the same underlying platform.
Extending the analysis to the full 131+~submission GitHub dataset~\cite{io500submission2025} would improve vendor diversity and statistical robustness.

\emph{Future work:} Further work should extend this analysis to the full GitHub dataset~\cite{io500submission2025} (131+~sites) for temporal trend analysis and more robust statistical conclusions.
Specific directions include longitudinal per-node efficiency tracking, process-level straggler modeling from CSV log files, and automated anomaly detection to flag unusual submissions.

\paragraph{Acknowledgments:}
We thank the IO500 community and the IO500 foundation (\href{https://io500.org}{io500.org}), and the VI4IO initiative for making the submission data publicly available.

\FloatBarrier
\bibliographystyle{splncs04}
\bibliography{references}

\end{document}